\documentclass[suppldata]{interact}

\usepackage{epstopdf}

\usepackage{natbib}
\bibpunct[, ]{(}{)}{;}{a}{}{,}
\renewcommand\bibfont{\fontsize{10}{12}\selectfont}

\theoremstyle{plain}

\theoremstyle{definition}

\theoremstyle{remark}

\usepackage{subcaption}
\usepackage{graphicx}
\usepackage{natbib} 
\usepackage{amssymb}
\usepackage{amsfonts}
\usepackage{amsmath,bm}
\usepackage{threeparttable} 
\usepackage{multirow}
\usepackage{color}
\usepackage{float}
\usepackage{caption}
\usepackage{setspace}
\usepackage{calrsfs}
\usepackage[hidelinks=true]{hyperref}
\usepackage[ruled]{algorithm2e}
\usepackage{fullwidth}
\usepackage[noend]{algpseudocode}
\newcommand{\R}{\mathbb{R}}
\newcommand{\G}{\mathcal{G}}
\newcommand{\N}{\mathcal{N}}
\newcommand{\E}{\mathcal{E}}
\newcommand{\I}{\mathcal{I}}
\newcommand{\Path}{\mathcal{P}}
\newcommand{\Hgraph}{\mathcal{H}}
\newcommand{\Origin}{\mathcal{O}}
\newcommand{\Destination}{\mathcal{D}}
\usepackage{tabulary}
\usepackage{tabularx}
\usepackage{dcolumn}
\newcolumntype{b}{X}
\newcolumntype{s}{>{\hsize=.5\hsize}X}

\begin{document}

\articletype{}

\title{Short-distance commuters in the smart city}

\author{
\name{Francisco Benita\textsuperscript{a}, Garvit Bansal\textsuperscript{a}, Georgios Piliouras\textsuperscript{b} and Bige Tun\c{c}er\textsuperscript{a} \thanks{CONTACT E-mail: \{francisco$\_$benita, garvit\_bansal, georgios, bige\_tuncer\}@sutd.edu.sg}}
\affil{\textsuperscript{a}Architecture and Sustainable Design, Singapore University of Technology and Design, Singapore; \textsuperscript{b}Engineering Systems and Design, Singapore University of Technology and Design, Singapore}
}

\maketitle

\begin{abstract}
This study models and examines commuter's preferences for short-distance transportation modes, namely: walking, taking a bus or riding a metro. It is used a unique dataset from a large-scale field experiment in Singapore that provides rich information about tens of thousands of commuters' behavior. In contrast to the standard approach, this work does not relay on survey data. Conversely, the chosen transportation modes are identified by processing raw data (latitude, longitude, timestamp). The approach of this work exploits the information generated by the smart transportation system in the city that make suitable the task of obtaining granular and nearly real-time data. Novel algorithms are proposed with the intention to generate proxies for walkability and public transport attributes. The empirical results of the case study suggest that commuters do no differentiate between public transport choices (bus and metro), therefore possible nested structures for the public transport modes are rejected. 
\end{abstract}


\section{Introduction}\label{sec: introduction}

Over the last years, more and more attention  is being paid to the analysis of factors contributing to the travel mode choice  
behavior for short-distance trips \citep{Mackett2003,Limtanakool2006,Guo2009,Vij2013}. 
On the one hand, the increase in commuter's mobility has played an important role in the spatial transformation in advanced urban economies where their smart transportation systems and the active commuting behavior of individuals, \emph{via} self-generated traffic information, provides real-time transit data. On the other hand, trip purpose is a key element of establishing the commuters behavior. The transport mode choice may differ, for example, between home-work, home-school or leisure trips. Due to the nature of the home-school trips, that account for the self-selection bias, there is an emerging body of literature examining the relationship between travel-to-school mode choice and a number of factors that might affect transportation mode choice \citep{Ewing2004,Mcdonald2008,Muller2008,Kamargianni2015,Ermagun2017}.  

The travel-to-school mode choice literature stress the preponderance of actively-related travel-modes like walking or biking due to short travel-to-school distances. Nevertheless, the classical scenarios for short-distance are: walk, public transport (bus, metro, tram, etc.) and car. The potential for switching short car trips to other modes has been addressed, among others, in \cite{Mackett2003} and \cite{Kim2008}. The empirical findings suggest that walk mode is less likely as commuters' age increases, and they are more likely to use car if they can drive or are accustomed to. Therefore, some works either restrict or control the study by commuters age group. Short trips not only provide opportunities for physical exercise for trip makers (students) but also has lead concerns at government level to the introduction of a wide range of policies to address the problems of high levels of car usage (students being driven). Road congestion, CO$_2$ emissions, air pollution, traffic accident or noise pollution are some of the undesired effects. Local governments through their Departments of Transport are attempting to encourage greater participation for walking, riding and using of public transport for short trips, e.g., under 20 minutes \citep{Mackett2003}. Moreover, in developing and some advanced urban economies, road-based public transport (bus/metro/tram and  pedestrian paths) is the only mean to access employment, education, and public services due to the limited access to private vehicles. In land-constrained countries such as Hong Kong, Japan, Singapore or South Korea, the Transportation Departments are encouraging a joint walking/public transport travel mode in place of car with special focus on promoting walking trips for short-distance trips among youths \citep{Haque2013}. 

In spite of the health benefits of walking/cycling over public transport options, such number of trips undertaken by youth commuters (students) is still low \citep{Panter2008}. The empirical evidence presented by \cite{Panter2008}, \cite{Guo2009}  and \cite{Haque2013} indicates that commuters place high value on urban amenities such as shorter commute time or neighborhood walkabiliy. This indicates that many residents want to live less automobile-dependent lifestyles, or even decrease the total number of public transport trips, if given suitable urban design features such as  walkable neighborhoods, traffic calming, street furniture, presence of traffic signals, crosswalks, among others.

Currently, a new body of literature that focuses on commuter's revealed preferences based on the automated  data collection  system of the smart cities has emerged. \cite{Schussler2008} infer trips and activities based on GPS data from three cities in Switzerland.  \cite{Asakura2012} investigate the behavior of passengers before and after changes to the train schedule. The authors perform train choice estimation models by using 3 million of records of smart card data of the Railway in Japan. In the case of Singapore, the smart city initiatives carried out by the government have been investigated by several studies. For instance, using  smart card data and/or GSM (cellular phone) data, the works in \cite{Holleczek2015}, \cite{Lee2014}, \cite{Sun2012} and \cite{Poonawala2016} explore the public (bus/metro) and private (taxi) transportation usage and transport mode preferences. \cite{Legara2018} perform an analysis of demographics of the commuting public using smart card data with more than 3 million of users. 
 
This study adopts the above-mentioned framework which exploits the  potential of automated data collection systems in smart cities. Using a unique dataset from the Singapore's National Science Experiment (NSE), this study presents a statistical multivariate research of the impact of several walking accessibility and public transport characteristics on the transportation mode choice for short-distance trips. The dataset is processed such that only walking and public transport (bus, metro) modes are analyzed.  By focusing on suitable walking trips which are less than 2.5km, the two main contributions of this work are: (i) to test whether or not youth commuters perceive differences between public transport modes, and; (ii) to identify the drivers that influence commuter's behavior to choose public transport over walking for short trips. To answer both research questions, there are introduced Binary Logit, Multinational Logit (MNL) and Nested Logit (NL) models to test the independence of irrelevant alternatives. The novelty of the study is the structure of the analyzed data and the suggested methodology that generates proxies for walkability and public transport attributes. 

In contrast to most previous studies, this work is not based on survey data. To the best of the authors' knowledge, this is the first non-survey-based work in address commuters' preferences for short-distance transportation modes. The NSE dataset consists of granular latitude/longitude row data  generated by daily behavior of tens of thousands of Singapore students that carry custom-made sensors for up to 4 consecutive days. Indeed, every 13 seconds, the sensor can accurately log its geographical location as well as other environmental factors such as relative temperature and humidity or noise levels \citep{Wilhelm2016}. The raw NSE data is processed using \emph{state-of-the-art} Machine Learning algorithms \citep{Monnot2016,Monnot2017,Wilhelm2017,Tunccer2017,Benita2019} to detect travel-to-school mode choices. 
 
Although the transportation mode detection algorithm allows to distinguish between, car, walk (with cycling mode embedded), taking bus and riding a metro; car trips were excluding due to the anonymity and confidentiality of commuters\footnote{For instance, if a student did not report car trips, then it is not possible to know whether he has access to car.}. To address walkability and public transport features, this work is built upon the advantages of the information generated by Singapore's smart transportation system. Nearly real-time attributes related to public transport mode were obtained from the API of the Land Transportation Authority of Singapore. The walkability attributes were constructed from Geographic Information System (GIS) data provided by Singapore's Urban Redevelopment Authority. The walkability proxies were derived by modeling the road street network as an undirected edge-weighted graph. In doing so, it is presented a detailed methodology that allows transportation authorities, urban planners and policy makers to measure preferences of commuters when making short trips.

The remainder of this article is organized as follows. After the introduction, Section \ref{sec: methodology} provides a theoretical and empirical discussion of possible representations of utility functions. Section \ref{sec: features} presents the determinants of mode choice as well as the detailed description of the proposed algorithms to compute proxies for public transport and walkability attributes. This is followed by the examination of the commuter's choice behavior in Section \ref{sec: results}. Finally, some concluding remarks can be found in Section \ref{sec: conclusions}.

\section{Theoretical and empirical framework}\label{sec: methodology}


Discrete choice models are based on the so-called random utility maximization. They assign a utility function $U_{ij}$ to each commuter $i \in I := \{1,\ldots,N\}$  given a set of (mutually exclusive) transportation modes $j \in J := \{1,\ldots,M \}$. The functions $U_{ij}$ are determined by a number of attributes related to both, the commuters $i \in I$, and the transportation modes $j \in J$. The utilities are represented by $U_{ij}=V_{ij} + \varepsilon_{ij}$,
where $V$ contains the deterministic part (observed) and $\varepsilon$ the stochastic part (unobserved). Let $c_i$ denote the alternative chosen by individual $i$, then the probability, $P_{ij}$, that commuter $i$ chooses transportation mode $j \in J$ is
\begin{equation}\label{probabilities}
\begin{aligned}
P_{ij} &= \text{Pr}(c_i = j) = \text{Pr}(U_{ij} > U_{ik} \quad \forall k\in J, k \neq 
j) \\
P_{ij} &= \text{Pr}(\varepsilon_{ik} - \varepsilon_{ij} \leq V_{ij} - V_{ik} \quad \forall k \in J, k \neq j),
\end{aligned}
\end{equation} 
where $P_{ij}$ will depend on the assumptions about the distribution of $\varepsilon$.

The observable part of the utility  function, $V$, must  be specified  to  be operational. Following the standard notation in the literature, one has \citep{Heiss2002}
\begin{equation}\label{determinsitic_part}
V_{ij}= \alpha_j + X^\top_{i}\beta_{j} + Y^\top_{ij}\gamma_j + Z_{ij}^\top \delta,
\end{equation}
where vector $X_i$ contains individual-specific variables related to each commuter $i\in I$, and they have alternative specific coefficients $\beta_j$. $Y_{ij}$ collects alternative-specific variables with alternative specific coefficients $\gamma_j$ whereas $Z_{ij}$ also collects alternative-specific variables, but with generic coefficient $\delta$. 

\begin{figure}[!h]
\centering 
  \begin{subfigure}[b]{0.39\textwidth}
    \includegraphics[width=\linewidth]{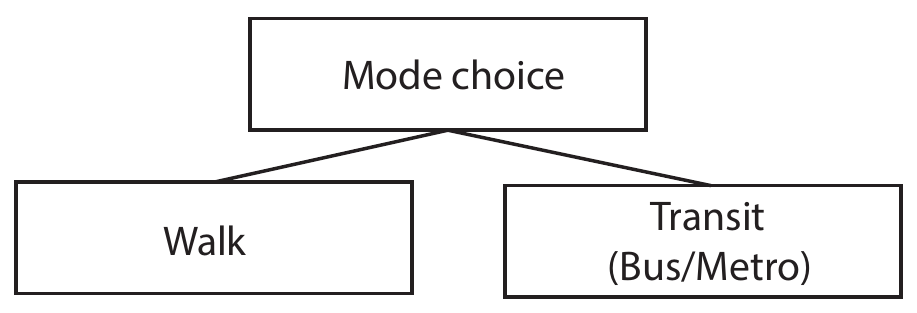}
    \caption{Binary Logit}\label{fig: logit}
  \end{subfigure}
  \begin{subfigure}[b]{0.57\textwidth}
    \includegraphics[width=\linewidth]{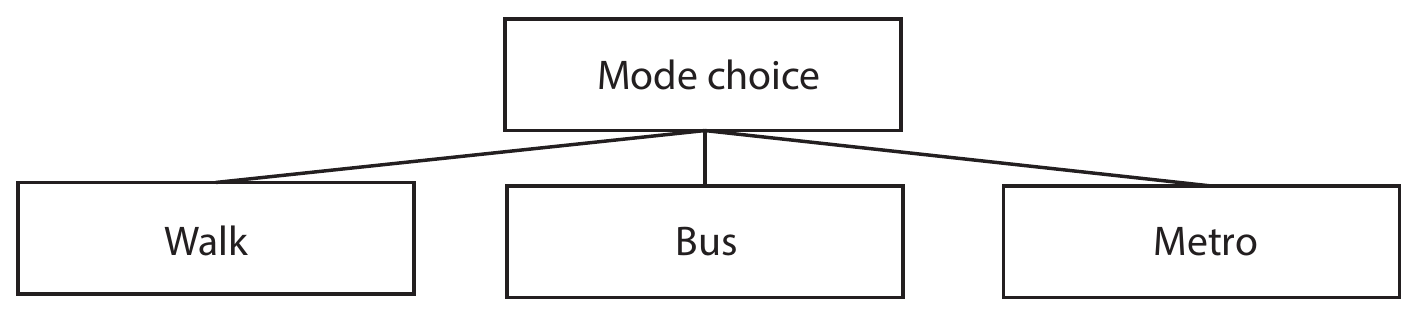}
    \caption{Multinomial Logit}\label{fig: multinomial}
  \end{subfigure}\\
   \begin{subfigure}[b]{0.39\textwidth}
    \includegraphics[width=\linewidth]{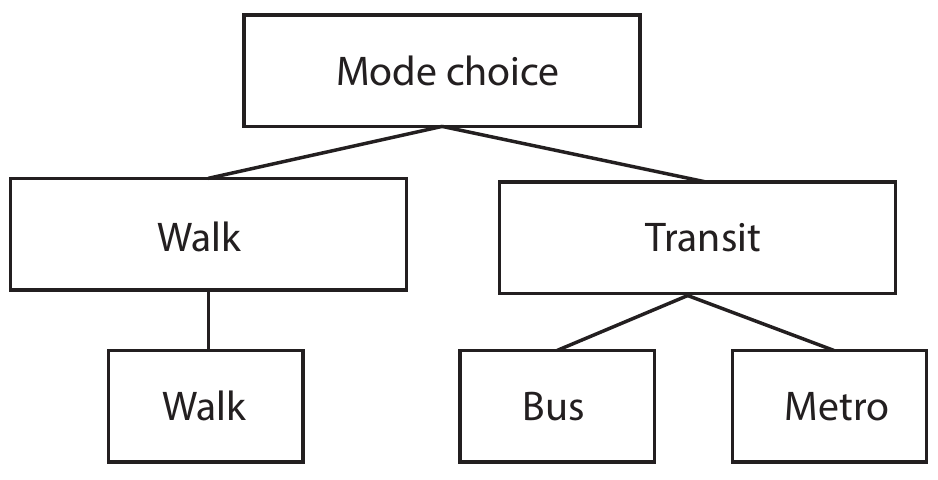}
    \caption{Nested Logit with degenerate branches}\label{fig: nested}
  \end{subfigure}
  \caption{Short-distance transport choice}\label{fig: trees}
\end{figure}

Since only utility differences are relevant for the choice according to equation \eqref{probabilities}, the parameters for one (the reference) transportation mode   must be normalized to zero for purposes of identification. Given the nature of the transportation modes, this study tests the structures shown in Figure  \ref{fig: trees}. First, the decision to walk or to use public transport is modeled under binary outcomes (Figure \ref{fig: logit}). Next, the commuting decision becomes more complex than the binary case and  the decision-maker $i\in I$ faces multiple choices: to walk, to use bus or to use metro (Figure \ref{fig: multinomial}). Finally, it is tested if in fact, public transport alternatives are nested as shown by Figure \ref{fig: nested}.

By comparing only two choices, between walking and using public transport, the commuter considers bus and metro options as perfect substitutes, this is, he would exchange one commute by bus for one commute by metro. At first glance, this may seem to be a rather strong assumption because bus lines have higher level of connectivity than metro lines. Specifically, decision-maker $i$ has two levels of utility, $U_{i,\text{walk}}$ and $U_{i,\text{transit}}$, that are associated with $y_{i}=0$ or $y_{i}=1$, respectively. Here, walking mode is the reference mode and the observable part of the utility function becomes $V_{ij} = \alpha + X^{\top}_i \beta_j$,
for $j \in J:=\{ 0,1 \}$. Note that this model depends only on individual attributes and 
the parameter $\alpha$ cannot vary across transportation modes. Thus, the binary logit model becomes  
\begin{equation}\label{logit_regression}\tag{Logit}
s_{i}=\alpha + X_i^\top\beta + \nu_i,
\end{equation}
where $s_i = U_{i,\text{transit}} - U_{i,\text{walk}}, \beta=\beta_{\text{transit}}-\beta_{\text{walk}}$, and $\nu_i ={\varepsilon_{i,\text{transit}}-\varepsilon_{i,\text{walk}}}$. 
 

By relaxing the assumption of perfect substitution between bus and metro, the first idea is to compare all three transportation modes (walk, bus and metro) within the same level of a three as shown in  Figure \ref{fig: multinomial}. Therefore, by choosing walking mode as reference mode one gets

\begin{equation}\label{multinomial}\tag{MNL}
\begin{aligned}
V^\prime_{ij}&=(V_{ij}-V_{i,\text{walk}})  \\
&=  (\alpha_j - \alpha_{\text{walk}}) + X_{i}^\top (\beta_j - \beta_{\text{walk}})  + (Y_{ij}^\top \gamma_{j} - Y_{i,\text{walk}}^\top \gamma_{\text{walk}}) + (Z_{ij}- Z_{i,\text{walk}})^\top \delta.
\end{aligned}
\end{equation}

Under the utility maximization principle, together with the assumption on the error terms $\varepsilon_{ij}$ follow a Gumbel distribution and  are distributed independently and identically across alternatives, the probability of commuter $i \in I$ choosing transportation mode $j \in J$, $P_{ij}$, is given by $P_{ij}=P_{i,\text{walk}} \cdot \text{e}^{V^\prime_{ij}}$. 



Next, the main assumption of utility functions stated in equation \eqref{multinomial} is the imposition of substitution patterns across transportation mode alternatives. This assumption is the so-called \emph{independent of irrelevant alternatives} (IIA) which states that when a commuter is asked to choose between walk, take bus or take metro, his odds of choosing, for example, walk over bus should not depend on whether metro alternative is present or absent. Under the IIA hypotheses, if relative walking time for bus decreases, the probability of choosing bus draws probability away from walk and taking metro  equally (in relative sense). The IIA property may not reflect the behavioral relationship among transportation mode choices. The idea is to allow public transport alternatives to be like each other in an unobserved way. This is, to have $\text{Cov}\{\varepsilon_{\text{walk}},\varepsilon_{j^\prime} \}\neq 0, j^\prime \in \{\text{bus,metro}\}$. Such nested representation is illustrated in Figure \ref{fig: nested}. First, commuter $i\in I$  decides between walk or using public transport, if public transport is chosen, then in a second stage, he chooses between bus and metro. Due to the nested alternative \emph{walk} only consist in a unique possible transportation mode  (i.e., to walk), one has degenerated branches.

The transportation modes are partitioned into two subsets, also called \emph{nests}, so that each alternative belongs to exactly one nest. Let $B_{\text{walk}}$ and $B_{\text{transit}}$ denote the two nests as represented by Figure \ref{fig: nested}. Hence, the probability of commuter $i \in I$ choosing transportation mode $j \in J := B_{\text{walk}} \cup B_{\text{transit}}$ equals $P_{ij}=\text{Pr}(c_i=j)=\text{Pr}\{ y_i=j| y_i \in B_k \} \cdot \text{Pr}\{ y_i \in B_k\}$
where $k\in \{\text{walk,transit} \}$. 
During the estimation procedure, for each nest $k$, it is estimated a dissimilarity parameter, $\lambda_k \in [0,1]$, measuring the degree of correlation of random shocks within each of the two types of transportation modes (walk and transit). However, due to $B_\text{walk}$ is a degenerated nest, one has $\lambda_{\text{walk}}=1$. Under the NL with degenerated branches model, the vector of unobserved utility $\varepsilon^\top=(\varepsilon_{i,\text{walk}}, \varepsilon_{i,\text{bus}}, \varepsilon_{i, \text{metro}})$ has generalized Extreme Value Type I distribution \citep{Hensher2005}. 
    
Hence, one can decompose the deterministic part of the utility function \eqref{determinsitic_part} into two parts such that 
\begin{equation}\label{decomposed_nested_formula}\tag{NL}
V_{ij}=  W_{ik}^\top \eta_{k}  + D_{ij}^\top \beta_j, \qquad j \in B_k, 
\end{equation}
where $W_{ik}$ depends only on variables that describe nest $k$, and $D_{ij}$ depends on variables that describe the transportation modes within the nest. Note that model  \eqref{decomposed_nested_formula} is fully general due to for any $W_{ik}$, one has $D_{ij}={V_{ij}-W_{ik}}$.  Without loss of generality, the alternative-specific intercept $\alpha_{j}$ and alternative-specific variables, $Z_{ij}$, with  generic coefficient $\delta$ can be easily allocated into equation \eqref{decomposed_nested_formula}. 

From the seminal work of \cite{Mcfadden1978}, the probability of commuter $i \in I$ to choose transportation mode  $j \in B_{\text{transit}}$ is given by
$P_{ij} = ( P_{iB_{\text{transit}}}) \cdot (P_{ij|B_{\text{transit}}})
$. Here $P_{iB_{\text{transit}}}$ is the marginal probability of choosing public transport mode in nest $B_{\text{transport}}$, and $P_{ij|B_{\text{transit}}}$ is the conditional probability of individual $i\in I$ choosing a transportation mode $j$ given that $j\in B_{\text{transit}}$. In this case, one has
\[
P_{ij}=
\dfrac{\text{e}^{( W^\top_{i, \text{transit}} \eta_{\text{transit}} + \lambda_\text{transit} \cdot IV_{i,\text{transit}})}}{\text{e}^{(W^\top_{i, \text{walk}}\eta_{\text{walk}} + V_{i,\text{walk}})} + 
\text{e}^{(\alpha_j + W^\top_{i, \text{transit} }\eta_{\text{transit}} + \lambda_\text{transit} \cdot IV_{i,\text{transit}})}} \cdot \frac{\text{e}^{\left(\frac{D_{ij}^\top \beta_j}{\lambda_{\text{transit}}}\right)}}{\text{e}^{\left( \frac{D_{i,\text{bus}^\top }\beta_{\text{bus}}}{\lambda_\text{transit} }\right)}\text{e}^{\left( \frac{D_{i,\text{metro}}^\top \beta_{\text{metro}}}{\lambda_\text{transit} }\right)}}
\]
where $IV_{i,\text{transit}}=\text{ln} \left( \text{e}^{\frac{D_{i,\text{bus}}^\top \beta_{\text{bus}}}{\lambda_\text{transit}}} +  \text{e}^{\frac{D_{i,\text{metro}}^\top \beta_{\text{metro}}}{\lambda_\text{transit}}} \right)  , IV_{i,\text{walk}}= V_{i,\text{walk}}$ are the \emph{inclusive values}. A similar algebraic expression can be derived to obtain the probability of commuter  $i \in I$ to walk by choosing $j \in B_{\text{walk}}$.

Because larger values of $\lambda_{\text{transport}}$ mean  greater independence and less correlation, therefore if $\lambda_{\text{transport}}=1$, it  means independence among all three transportation modes, which reduces the NL model MNL model. 
The inclusive value terms enter as  explanatory variables in the NL model, However, since it is imposed the constraint $\lambda_{\text{walk}}=1$, only the coefficient $\lambda_{\text{transit}}$ is estimated. To be consistent with the random utility model, it should be tested that $\lambda_{\text{transit}}\in [0,1]$.

\section{Path walkability and public transport infrastructure}\label{sec: features}


\subsection{The National Science Experiment dataset}

The Singapore's National Science Experiment (NSE) is the first nationwide project with the main goal of attract students and graduates to study and work in the science, technology, engineering and math (STEM) sectors. NSE is part of the Smart Nation Programme Office, 2015 initiative, involving over 100,000 students from over 200 schools from primary (ages 7-12), secondary (13 to 16) and junior colleges (17 and 18). The students wore a sensor, called SENSg, for one week in a rotated schema during 2015 and 2016. The SENSg device was designed to collect ambient temperature, relative humidity, atmospheric pressure, light intensity, sound pressure level, and 9-degree of freedom motion data. The sample rate of the SENSg device is about 13-15 seconds and the raw data is about 50 million lines for each NSE year. Using data from NSE 2016 (about 50,000 students and 100 schools) it is possible to observe the commuting behavior among youth Singaporeans. During the data collection, students did not actively report any kind of information. All explanatory variables that enter the utility functions were indirectly obtained by processing the information generated by the SENSg device as well as other official sources of information. The key variable, the chosen transportation mode $j \in J$ was obtained by implementing the Machine Learning algorithms described in \cite{Wilhelm2017}. 
Morning trips, from home to school were extracted, and a total of 2,959 trips (made by 1,832 different students) were considered for the analysis. The number of trips/students was dramatically reduced from the original data due to the following filters: (i) morning trips only; (ii) exclude car trips; (iii) short-distance trips with an observed length of less than 2.5km only; and (iv) exclude trips whose corresponding origin-destination request in Google Directions API gave empty results. 

\subsection{Alternative-specific variables with generic coefficient}
The value of travel time refers to the cost of time spent on transport. For short-distance trips, one can assume that relative walking time or relative walking distance are important  factors that influence the choice of mode of transport. 
Relative walking distance was also used in \cite{Tan2018} to evaluate the effects of active morning commutes on students' overall daily walking activity in Singapore. Using the algorithms described in \cite{Monnot2016}, students' homes
and schools' locations were obtained in latitude/longitude formats. Let $\Origin$ and $\Destination$ represent the set of origins and destinations, respectively. Then, it is worth to mention that destinations $d^i \in \Destination$ were calculated as the exact ending point commuters needed to reach. For example, if students $i,i^\prime \in I$ attend the same school, it is most likely that their exact destination points $d^i,d^{i^\prime} \in \Destination$  may differ from each other. This is not a trivial issue due to in a 2.5 km distance, the decision whether to chose transportation mode $j\in J$ may be substantially affected. For every trip in the dataset, one has an origin $o^i \in  \Origin$ and a destination $d^i \in  \Destination$ that may be repeated in the sample as long as student $i$ appears more than once in the dataset, i.e., home locations do not usually change for a student $i$. Every $(o^i,d^i)$ pair is indexed with a departure time $\tau^i$ that differs among days for every student $i$. For every 3-tuple $(o^i,d^i,\tau^i)$, the walking distance (if $j=\text{walk}$) was obtained using Google Directions API querying according to the exact observed departure time $\tau^i$. In the case of public transport modes, the Land Transportation Authority of Singapore API was used. 

\subsection{Individual-specific attributes}

The NSE 2016 data allows to have access to detailed information about student's mobility, however privacy and security concerns appear. This study meets the requirements of the Singapore's Personal Data Protection Act (PDPA) by using anonymous trips. Lucky, it is possible to control by individual heterogeneity using indirect measures. Student type (primary-, secondary- or junior college students) is included in the model by entering binary variables. Student type attributes controls for unobserved individual heterogeneity among commuters. For instance, a reasonable assumption is that primary school students are more likely to walk due to the Ministry of Education ensures that the enrollment to primary schools is based on home-school distance. On the other hand, time is one of the larger costs of transportation. Thus, commuters can reduce their relative walking time by choosing public transport mode but being charged distance-based fares. It is not too restrictive to assume that commuters with higher income levels are more willing to pay transit fares due to their lower marginal cost paid if they choose public transport modes. 
The average residential rental price (at Subzone\footnote{The Urban Redevelopment Authority divides Singapore into 5 Regions, 55 Planning Areas and 323 Subzones.} level), as a proxy for income, is introduced to address this attribute, see \cite{Benita2019}.

Before discussing transportation mode-specific attributes, it should be noticed that several of these attributes may be correlated with population size factors, embedded at the $(o^i,d^i)$ pairs. In urban settings, population density, either at the origin or destination location, is correlated with congested road networks, longer peak times, higher walking or transit accessibility \citep{Guo2009}. Using data from the  General Household Survey 2015 elaborated by the Department of Statistics Singapore, this study includes population density (persons per km$^2$) related to the home-location ($o^i$) to control for such factors at Subzone level.

\subsection{Walk-specific attributes}

The mix of commercial facilities, residential areas, road facilities, parks or open spaces has enabled commuters to walk  short-trips. The theory is that population density makes walking efficient by creating demand for destinations and decreasing the appeal of alternative transportation modes \citep{Owen2004}. By processing the land use shapefile of the Singapore Master Plan 2014 Use of Land, a mixed land uses index is computed for every Subzone. The measure, called entropy index, combines multiple walkable destinations. The index is computed as suggested by the seminal work of \cite{Frank2005}, and considers four types of land use: (i) road segments; (ii) commercial areas; (iii) residential areas; and (iv) parks and open spaces. The entropy index takes values between 0 and 1, with 0 representing homogeneity (all land uses are of a single type), and 1 representing heterogeneity. 

In the same vein, researchers from GIScience, city planning, and urban design have been developed multiple measures of walkability \citep{Frank2010}. This work made use of the road network shapefile provided by Singapore's Land Transportation Authority to compute several proxies of walkability. First, it is introduced a new metric that accounts for diversity of the walking paths. The basic idea is to consider the road network as an undirected edge-weighted graph, then the shortest path distance for every ($o^i,d^i$) pairs is computed using Dijkstra's shortest path algorithm. Next, one can introduce a relaxation ($\epsilon>1$) to allow commuter $i \in I$ to find how many different walking paths exist that are $\epsilon$ times longer that the shortest path. 
The details of the $\epsilon-$shortest path algorithm are given in Table \ref{tab: algorithm epsilon shortest path} in the Appendix. 

Lastly, there are used load centrality and degree centrality as proxies for connectivity. Load centrality and degree centrality are computed for both,  $o^i$  and $d^i$. Every home and school location is assigned to the nearest node in the network and then both metrics are computed from the network graph using Python's NetworkX \citep{Hagberg2008}. Recall that load centrality of a node is the fraction of all shortest paths that pass through that node whereas degree centrality for a node is the fraction of nodes it is connected to. 

\subsection{Public transport-specific attributes}

The effect of departing at peak hours is investigated.  A dummy variable is included taking the value of 1 if commuter $i \in I$ departed during AM peak periods and 0 otherwise. The classical approach is to assume some time interval, for example between 6am and 9am, to take the value of 1. Nevertheless, the smart transportation system provided by the API of the Land Transport Authority of Singapore allows to inform commuters nearly-real-time peak times. To obtain more granular data related to each commuter, the computed peak binary variable will depend on his own departure time as well as his own route if he decides to chose public transport modes. 

The effect of transit fares on transportation mode choice have been widely studied in the literature. Nevertheless, the data used in this work does not allow to include large variations on public transport fares. 
For instance, in Singapore, if a commuter $i \in I$ pays using his student ID card, the cost for a trip up to 3.2 km is 37 Singapore cents (about  28 US cents). Longer trips may be rather unusual for students that live at most 2.5km home-school walking distance. With an attempt to get weak statistical significance in the models, transit fare price is introduced in the utility function using data from the Land Transportation Authority. Next, accessibility and connectivity to public transport is proxied by an index that measures the proportion of Subzones commuter $i\in I$  can reach from his Subzone of origin within a single-trip without transfers (by either bus or metro). 
The details of the procedure  can be found in Table \ref{tab: accessibility} in the Appendix. Finally, service frequency affects commuters' utility function. In  Singapore, metro and bus lines operate at double frequency during morning and evening peaks. However, waiting times are still unbalanced during off-peak time \citep{Sun2014}. By using the Singapore's Land Transportation Authority API, every 3-tuple $(o^i,d^i,\tau^i)$ is appended with its respective frequency (in minutes). 

\section{Results and discussion}\label{sec: results}

\subsection{Descriptive analysis}

The final dataset contains information from 2,959 trips made by 1,832 different commuters. The summary statistics are shown in Table \ref{tab: summary statistics}. The walking trips account for  22.4\% of the total  whereas 77.6\% (72\%  bus and only 5.6\% metro) of the trips were made by public transport. If commuters decide to walk, the average walking time is 24 minutes with a small standard deviation of only 4.5 minutes. Alternatively, if commuters decide to use public transport options, the average relative walking time reduces to only 9.3 minutes. 

About 82.1\% of the commuters are young students from primary and secondary  school, and only 17.8\% are commuters from junior college. The unbalanced nature of the dataset can be explained due to older commuters choose driving or being driven as transportation modes, hence excluded from this work. 
It should be highlighted that frequency (in minutes) of the public transport modes is very high.  If the commuter misses a bus/metro, he only needs to wait, on average, 3.4 minutes to take the next one. 
In addition, the commuters are exposed to relatively heterogeneous mixes of land use (entropy). Lastly, the $\epsilon-$shortest paths algorithm revels that the average commuter has access to 959 different walking paths that are no more than 10\% longer that his corresponding shortest path. Due to computational time constraints, the $\epsilon-$shortest paths algorithm was decided to stop after computing 3,000 different walking paths. 

\begin{table}[!h]
\centering
\caption{Summary statistics}
\label{tab: summary statistics}
\begin{small}
\begin{tabular}{rrrrrrr}
\hline \hline \\[-1.8ex] 
\multicolumn{1}{c}{Variable}                   & \multicolumn{1}{c}{Frequency} & \multicolumn{1}{c}{Percentage} & \multicolumn{1}{c}{Mean} & \multicolumn{1}{c}{Std. Dev.} & \multicolumn{1}{c}{Min} & \multicolumn{1}{c}{Max} \\ \hline
Mode walk                                      & 663                           & 22.4\%                         &                          &                               &                         &                         \\
Mode bus                                       & 2,131                         & 72.0\%                         &                          &                               &                         &                         \\
Mode metro                                     & 165                           & 5.6\%                          &                          &                               &                         &                         \\
Rel. walking distance$^*$ (mode walk)          &                               &                                & 2.0                      & 0.3                           & 0.82                   & 2.5                     \\
Rel. walking distance$^*$  (mode transit)      &                               &                                & 0.7                      & 0.3                           & 0                       & 2.1                     \\
Rel. walking time$^{**}$ (mode  walk)          &                               &                                & 24.8                     & 4.5                           & 4.7                     & 32.8                    \\
Rel. walking time$^{**}$ (mode  transit)       &                               &                                & 9.3                      & 4.5                           & 0.0                     & 28.2                    \\
Primary School                                 & 723                         & 39.5\%                         &                          &                               &                         &                         \\
Secondary School                               & 782                        & 42.7\%                         &                          &                               &                         &                         \\
Junior College                                 & 327                           & 17.8\%                         &                          &                               &                         &                         \\
Rental price (SGD)                             &                               &                                & 478.5                    & 171                           & 318                     & 1,799                   \\
Population density                             &                               &                                & 27,137                   & 11,373                        & 11                      & 48,557                  \\
\multicolumn{1}{l}{\textit{Transit variables}} &                               &                                &                          &                               &                         &                         \\
Peak                                           & 1,070                         & 36.2\%                         &                          &                               &                         &                         \\
Transit fare (cents)                           &                               &                                & 39.4                     & 4.9                           & 37                      & 58                      \\
Accessibility (transit)                        &                               &                                & 124                      & 45                            & 27                      & 231                     \\
Frequency$^{**}$                              &                               &                                & 3.4                      & 5.0                           & 4                       & 24                      \\
\multicolumn{1}{l}{\textit{Walking variables}} &                               &                                &                          &                               &                         &                         \\
Entropy                                        &                               &                                & 0.54                     & 0.21                          & 0                       & 1                       \\
Accessibility (walking)                        &                               &                                & 959                      & 1,233                         & 1                       & 3,000                   \\
Load centrality (Home)                         &                               &                                & 0.007                    & 0.030                         & 0                     & 0.154                   \\
Load centrality (School)                       &                               &                                & 0.006                    & 0.029                         & 0                     & 0.141                   \\
Degree centrality (Home)                       &                               &                                & 0.004                    & 0.026                         & 0                      & 0.013                   \\
Degree centrality (School)                     &                               &                                & 0.004                    & 0.026                         & 0                      & 0.013                  
\\
\hline \hline
\end{tabular}
   \begin{tablenotes}
      \small
      \item \textit{Note:} $^*$ in kilometers, $^{**}$ in minutes.
    \end{tablenotes}
\end{small}
\end{table}

\begin{figure}[!h]
\centering 
\caption{Road street network and streets usage}
    \includegraphics[scale=0.3]{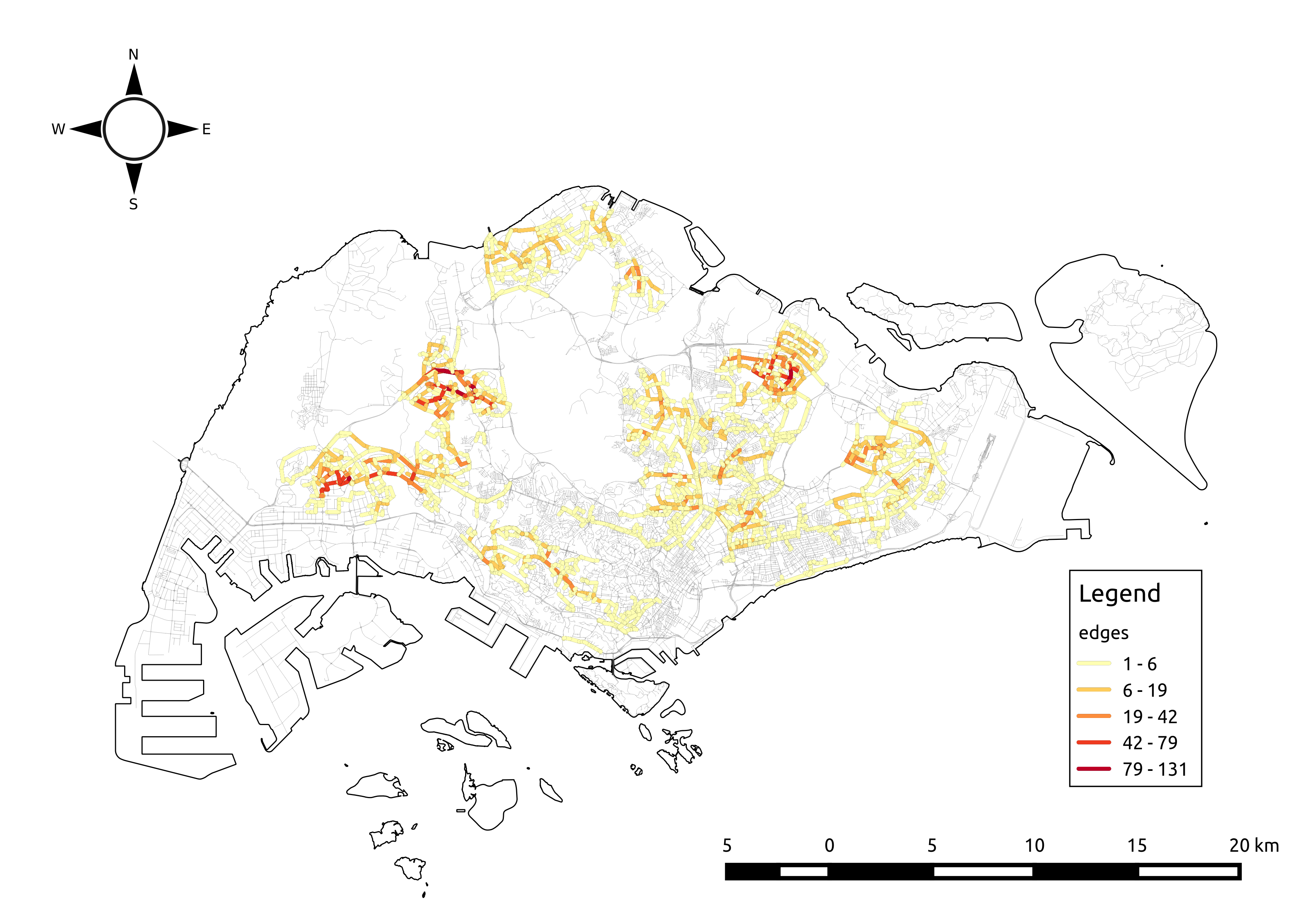}
\caption*{ \textit{Note:} Roads usage clustered by the Jenks natural breaks optimization algorithm.}
 \label{fig: map of Singapore}
\end{figure}

Figure \ref{fig: map of Singapore} displays the road network of Singapore. After processing the  road street network shapefile according to the algorithm presented in Table \ref{tab: algorithm cleaning graph} (see the Appendix), every $(o^i,d^i)$ pair was mapped to its corresponding path. Then, without looking at the chosen transportation mode, all paths were  plotted weighting those road segments (edges) according to it's frequency flow. Darker colors in Figure \ref{fig: map of Singapore} correspond to more visited segment roads. From the figure, it is possible to observe that the commuting area covered by NSE 2016 data is well dispersed throughout the city. 

\subsection{The binary choice case}

The results from equation \eqref{logit_regression} are shown in Table \ref{tab: logit} and four specifications are presented. The basic model in column (1), excludes transit and walking attributes. This model includes only relative walking distance/time, controls for school type, proxy for household income and population density. The augmented model in column (2) adds attributes related to public transport mode such as departing at peak time, transit fare, accessibility and frequency. The specification in column (3) is an extended model that includes walkability attributes. Finally model 4 excludes the main explanatory variables ``relative walking distance/time'' as well as individual controls with the attempt to test the quality of the model under only transit and walking variables.  The effect of relative walking distance and relative walking time seems to be not consistent in signs due to potential multicollinearity. In fact, one has Corr(Walking distance$_{\text{walk}}$, Walking time$_{\text{walk}}$)=0.976 and  Corr(Walking distance$_{\text{transit}}$, Walking time$_{\text{transit}}$)=0.996. Therefore, it would be possible to omit one of the variables without losing much explanatory power.

\begin{table}[h!] \centering 
  \caption{Parameter estimates for mode choice, walking vs transit} 
  \label{tab: logit} 
\begin{footnotesize}
\begin{tabular}{@{\extracolsep{5pt}}lD{.}{.}{-3} D{.}{.}{-3} D{.}{.}{-3} D{.}{.}{-3} } 
\\[-1.8ex]\hline 
\hline \\[-1.8ex] 
 & \multicolumn{4}{c}{\textit{Dependent variable:}} \\ 
\cline{2-5} 
\\[-1.8ex] & \multicolumn{4}{c}{Walk (0) vs Transit (1)} \\ 
 & \multicolumn{1}{c}{Basic} & \multicolumn{1}{c}{Augmented} & \multicolumn{1}{c}{Extended} & \multicolumn{1}{c}{NA} \\ 
\\[-1.8ex] & \multicolumn{1}{c}{(1)} & \multicolumn{1}{c}{(2)} & \multicolumn{1}{c}{(3)} & \multicolumn{1}{c}{(4)}\\ 
\hline \\[-1.8ex] 
 Rel. walking distance (Walking) & 3.207^{***} & 3.251^{***} & 2.972^{***} &  \\ 
  & (0.609) & (0.611) & (0.586) &  \\ 
  Rel. walking distance (Transit) & -5.857^{***} & -4.883^{***} & -3.943^{**} &  \\ 
  & (1.711) & (1.824) & (1.876) &  \\ 
  Rel. walking time (Walking) & 0.012 & -0.013 & 0.013 &  \\ 
  & (0.047) & (0.047) & (0.045) &  \\ 
  Rel walking time (Transit) & 0.391^{***} & 0.304^{**} & 0.229 &  \\ 
  & (0.132) & (0.142) & (0.146) &  \\ 
  Rental price (log) & -0.383^{*} & -0.477^{*} & -0.878^{***} &  \\ 
  & (0.215) & (0.254) & (0.273) &  \\ 
  Pop. density (log) & -0.064 & -0.091 & -0.095 &  \\ 
  & (0.066) & (0.071) & (0.073) &  \\ 
  Primary school & 0.347^{**} & 0.298^{*} & 0.330^{*} &  \\ 
  & (0.159) & (0.166) & (0.170) &  \\ 
  Secondary school & 0.699^{***} & 0.718^{***} & 0.728^{***} &  \\ 
  & (0.160) & (0.165) & (0.169) &  \\ 
  Peak &  & -0.346^{***} & -0.355^{***} & -0.404^{***} \\ 
  &  & (0.106) & (0.107) & (0.096) \\ 
  Transit fare &  & 72.511 & 72.376 & 70.682 \\ 
  &  & (1,530.793) & (1,520.157) & (968.995) \\ 
  Accessibility (Transit) (log) &  & 0.122 & 0.049 & -0.233 \\ 
  &  & (0.162) & (0.176) & (0.142) \\ 
  Frequency &  & -0.047^{**} & -0.035 & 0.072^{***} \\ 
  &  & (0.021) & (0.022) & (0.020) \\ 
  Frequency (NA) &  & -0.480^{**} & -0.402 & 0.729^{***} \\ 
  &  & (0.244) & (0.252) & (0.221) \\ 
  Entropy &  &  & 0.665^{***} & 0.416^{*} \\ 
  &  &  & (0.254) & (0.226) \\ 
  Accessibility (Walking) &  &  & 0.019 & 0.126^{***} \\ 
  &  &  & (0.020) & (0.017) \\ 
  Load Centrality (Home) &  &  & 7.430^{**} & 5.425^{*} \\ 
  &  &  & (3.524) & (3.259) \\ 
  Load Centrality (School) &  &  & -1.693 & -9.092^{***} \\ 
  &  &  & (3.761) & (3.438) \\ 
  Degree Centrality (Home) &  &  & -67.149^{*} & -55.388^{*} \\ 
  &  &  & (38.018) & (32.982) \\ 
  Degree Centrality (School) &  &  & -43.231 & -27.437 \\ 
  &  &  & (33.443) & (30.253) \\ 
  Constant & -1.490 & 71.748 & 74.339 & 71.595 \\ 
  & (1.616) & (1,521.996) & (1,511.421) & (963.426) \\ 
   \hline \\[-1.8ex] 
Observations & \multicolumn{1}{c}{2,959} & \multicolumn{1}{c}{2,959} & \multicolumn{1}{c}{2,959} & \multicolumn{1}{c}{2,959} \\ 
Log Likelihood & \multicolumn{1}{c}{-1,232.754} & \multicolumn{1}{c}{-1,142.070} & \multicolumn{1}{c}{-1,130.384} & \multicolumn{1}{c}{-1,336.365} \\ 
Akaike Inf. Crit. & \multicolumn{1}{c}{2,483.508} & \multicolumn{1}{c}{2,312.140} & \multicolumn{1}{c}{2,300.769} & \multicolumn{1}{c}{2,696.730} \\ 
\hline 
\hline \\[-1.8ex] 
\textit{Note:}  & \multicolumn{4}{r}{$^{*}$p$<$0.1; $^{**}$p$<$0.05; $^{***}$p$<$0.01} \\ 
\end{tabular} 
\end{footnotesize}
\end{table} 

An increase in relative walking distance in the public transport mode leads to a decrease in the probability of choosing transit. This is a well-known result detailed in \cite{Ermagun2017}.  Surprisingly, the average rental price has unexpected sign. In the extended model (3), an increase in the rental price leads to a decrease in the probability of choosing public transport at 1\% significance level. This is, if the proxy is correct, commuters from households with higher income tend to walk more compared to commuters from households with lower level of income. As suspected, transit fare has no effect (at least at the 10\% significance level) due to the short-distances traveled and the homogeneity, see  Table \ref{tab: summary statistics}. 
School type dummy variables are statistically significant under all three specifications. In particular, the younger commuters form primary and secondary school students tend to use more often public transport modes compared to the junior college commuters. The finding is contrary to the assumption that younger students walk more often due to the priority admission to schools based on distance. This result is in line with the empirical evidence provided in \cite{Tan2018}. 

Departing during peak hours decreases the probability of choosing public transport. In the extended model (3), the odds of taking public transport are estimated to be 1.43 ($\approx\frac{1}{\text{e}^{-0.355}}$) times as high for commuters who do depart off-peak as for commuters who depart during peak hours. The proxy for public transport accessibility does not has any statistical effect. However, the proxy for walking accessibility computed by the $\epsilon-$shortest paths algorithm is statistically significant under model (4). It was expected to obtain a negative relationship between frequency and public transport mode. This relationship is clear under models (2) and (3), but a change in sing is reported under model NA. The variable frequency (NA) is a dummy variable that accounts for those trips where frequency was not able to obtain with the Singapore's Land Transportation Authority API. Load centrality and (normalized) degree centrality is robust only at the origin node, i.e., home location. For example, an increase in road street network connectivity increases the probability of commuters to walk. The finding similar to the empirical evidence in \cite{Panter2008}, \cite{Guo2009}  and \cite{Haque2013}. 

One way to compare all four models in Table \ref{tab: logit} is by  means of their Akaike Information Criteria (AIC) score. Model (3) reports the lowest AIC which suggests a good fit compared to the rest of the specifications. By using the likelihood ratio test, based on the Log Likelihood information, one can compare the basic model (1) with the extended model (3). Simply calculations yield  LR(Model 1, Model 3)=$-2 (-1,232.754 - (-1,130.884))=203.74$. Comparing with the critical value of $\chi_{20-9,0.05}^2=19.67$ (at the 5\% confidence level) the decision is to reject the null hypothesis and conclude that there is evidence that public transport attributes and walkability attributes have differential effects on the chosen transportation mode.

\subsection{The differences between public transport modes}

It follows to test the hypotheses of differences between public transport modes according to figures \ref{fig: multinomial} and  \ref{fig: nested}. The \ref{multinomial} model can be considered a special case of a \ref{decomposed_nested_formula} model when the three alternatives are found to be uncorrelated and they can be treated as three parallel choices in the same nest. Table \ref{tab: mnl and nl} displays the main results. To avoid possible issues related to multicollinearity, as found in Table \ref{tab: logit}, relative walking time is excluded from the models. Relative walking distance enters the models as alternative-specific variable with generic coefficient (i.e., vector $Z_{ij}$ in equation \eqref{determinsitic_part}). This improves the efficiency of the estimates and allows to obtain an interpretation of the coefficient as the implicit value of distance in terms of utility. For simplicity, the remaining explanatory variables enter the models as individual-specific features with alternative-specific coefficients (i.e., vector $X_{i}$ in equation \eqref{determinsitic_part}). To take advantage form previous findings, only the robust variables in Table \ref{tab: logit} are included in the \ref{multinomial} and \ref{decomposed_nested_formula} models. 

\begin{table}[!h] \centering 
  \caption{Parameter estimates for mode choice, walking vs transit vs metro} 
  \label{tab: mnl and nl} 
\begin{scriptsize}
 \begin{tabular}{@{\extracolsep{5pt}}lD{.}{.}{-3} D{.}{.}{-3} D{.}{.}{-3} D{.}{.}{-3} } 
\\[-1.8ex]\hline 
\hline \\[-1.8ex] 
 & \multicolumn{4}{c}{\textit{Dependent variable:}} \\ 
\cline{2-5} 
\\[-1.8ex] & \multicolumn{4}{c}{Walk vs Transit (Bus and Metro)} \\ 
 & \multicolumn{1}{c}{MNL (basic)} & \multicolumn{1}{c}{MNL (augmented)} & \multicolumn{1}{c}{NL (basic)} & \multicolumn{1}{c}{NL (augmented)} \\ 
\\[-1.8ex] & \multicolumn{1}{c}{(1)} & \multicolumn{1}{c}{(2)} & \multicolumn{1}{c}{(3)} & \multicolumn{1}{c}{(4)}\\ 
\hline \\[-1.8ex] 
 Intercept$\times$bus & -0.560 & 0.207 & -0.207 & 0.358 \\ 
  & (1.181) & (1.217) & (0.974) & (1.066) \\ 
  \hspace{29pt}          $\times$metro  & -9.440^{***} & -8.350^{***} & -8.412^{***} & -7.696^{***} \\ 
  & (2.060) & (2.123) & (2.085) & (2.142) \\ 
  Rel. walking distance& -1.852^{***} & -1.814^{***} & -1.592^{***} & -1.602^{***} \\ 
  & (0.090) & (0.092) & (0.144) & (0.146) \\ 
  Rental price  (log)$\times$bus & -0.059 & -0.213 & -0.112 & -0.233 \\ 
  & (0.194) & (0.199) & (0.158) & (0.171) \\ 
  \hspace{59pt}                    $\times$metro  & 1.178^{***} & 0.982^{***} & 1.005^{***} & 0.869^{**} \\ 
  & (0.333) & (0.340) & (0.340) & (0.347) \\ 
  Peak$\times$bus &  & -0.610^{***} &  & -0.498^{***} \\ 
  &  & (0.103) &  & (0.100) \\ 
  \hspace{14pt}                    $\times$metro  &  & -0.497^{***} &  & -0.406^{**} \\ 
  &  & (0.193) &  & (0.198) \\ 
  Entropy$\times$bus &  & 0.799^{***} &  & 0.699^{***} \\ 
  &  & (0.222) &  & (0.192) \\ 
 \hspace{25pt}                    $\times$metro &  & 0.237 &  & 0.215 \\ 
  &  & (0.417) &  & (0.398) \\ 
  Accessibility (Walk)$\times$bus &  & 0.0002^{***} &  & 0.0002^{***} \\ 
  &  & (0.00005) &  & (0.00005) \\ 
  \hspace{68pt}                    $\times$metro  &  & 0.0004^{***} &  & 0.0003^{***} \\ 
  &  & (0.0001) &  & (0.0001) \\ 
  Load Centrality (Home)$\times$bus &  & 7.539^{***} &  & 6.378^{**} \\ 
  &  & (2.834) &  & (2.818) \\ 
  \hspace{82pt}                    $\times$metro  &  & 3.767 &  & 3.037 \\ 
  &  & (5.282) &  & (6.073) \\ 
  Load Centrality (School)$\times$bus &  & -6.690^{**} &  & -5.415^{**} \\ 
  &  & (2.879) &  & (2.649) \\ 
  \hspace{84pt}                    $\times$metro  &  & -6.268 &  & -5.297 \\ 
  &  & (5.368) &  & (6.115) \\ 
  $\lambda_{\text{transit}}$ &  &  & 1.283^{***} & 1.235^{***} \\ 
  &  &  & (0.141) & (0.141) \\ 
School type controls & \multicolumn{1}{c}{Yes}& \multicolumn{1}{c}{Yes} & \multicolumn{1}{c}{Yes} & \multicolumn{1}{c}{Yes}\\ 
 \hline \\[-1.8ex] 
Observations & \multicolumn{1}{c}{2,959} & \multicolumn{1}{c}{2,959} & \multicolumn{1}{c}{2,959} & \multicolumn{1}{c}{2,959} \\ 
 McFadden R$^{2}$ & \multicolumn{1}{c}{0.144} & \multicolumn{1}{c}{0.166} & \multicolumn{1}{c}{0.145} & \multicolumn{1}{c}{0.168} \\ 
Log Likelihood & \multicolumn{1}{c}{-1,855.930} & \multicolumn{1}{c}{-1,806.846} & \multicolumn{1}{c}{-1,852.228} & \multicolumn{1}{c}{-1,804.455} \\ 
LR Test (basic vs augmented) & \multicolumn{1}{c}{} & \multicolumn{1}{c}{98.17$^{***}$ (Reject)} & \multicolumn{1}{c}{} & \multicolumn{1}{c}{95.55$^{***}$ (Reject)} \\
\hline 
\hline \\[-1.8ex] 
\textit{Note:}  & \multicolumn{4}{r}{$^{*}$p$<$0.1; $^{**}$p$<$0.05; $^{***}$p$<$0.01} \\ 
\end{tabular}
 \end{scriptsize} 
\end{table} 

It is worth noting that signs and magnitude of the estimated parameters are similar under all four specification. As expected, relative walking distance has a large negative effect in all four models. The coefficient is significant at the 1\% level which suggests that the distance walked is associated with a disutility, and the probability of choosing any transportation mode decreases as relative walking distance increases. The peak parameters for bus and metro are very similar. Keeping all other variables constant, if a commuter departs during peak hours,  the odds to choose public transport modes are about 40-45\% lower as compared to walk. The results for entropy are in line with those found in Table \ref{tab: logit}. This is, an increase in the diversity or walkable land uses, increases the probability of choosing public transport modes as compared to walk. The finding suggests that land use mix related more to public transport trips rather than walking trips. The (unexpected) sign of the entropy index coefficient can be explained by the fact that one of its components accounts for the road street area. Then, the entropy index includes information for both, pedestrian footpaths and roads for vehicles usage (correlated with public transport modes).

An increase on the walking accessibility, measured as the number of shortest paths in a 10\% gap, is correlated with the use of public transport modes. Despite the significance is at the 1\% level under the \ref{multinomial} and \ref{decomposed_nested_formula} models, the magnitude is nearly zero. What can be inferred from tables \ref{tab: logit} and \ref{tab: mnl and nl} is that the proxy for connectivity between origin and destination as well as the entropy index are closely related to accessibility to public transport modes. A potential explanation for the unexpected result is that the utility functions does not include \emph{real} walking paths that commuters may choose in a real-life scenario. In fact, it is quite common that commuters may cross through residential properties to take shortcuts. Yet, these paths are infeasible for both, the $\epsilon-$shortest path algorithm and the Google Directions API. The home-load centrality is significant for the bus mode, implying that higher connectivity to home increase the probability to use bus compared to walk. 

The likelihood-ratio statistic basic \emph{vs} augmented (LR Test (Basic, Augmented)) compares the basic model against the extended model under the same decision-making process. For example, in both cases, it is rejected the null hypothesis, indicating that the augmented models fit significantly better than the basic models. It is very important to observe the magnitude and statistical significance of $\lambda_{\text{transit}}$ in both \ref{decomposed_nested_formula} models. The dissimilarity parameter of \ref{decomposed_nested_formula} (basic) and \ref{decomposed_nested_formula} (augmented) are 1.28 and 1.23, respectively, with a statistical significance of 1\%. This provides statistical evidence supporting the idea that the dissimilarity parameter is greater than one, hence  the correlation between ``bus'' and ``metro'' is not statistically significant. Therefore, one can concludes that the \ref{multinomial} models are preferred over \ref{decomposed_nested_formula} specifications. This  main finding implies that, in Singapore, commuters do not differentiate between bus and metro, hence, both public transportation modes are seen as perfect substitutes.

\subsection{Test for individual effects}

The nature of the dataset allows to explore  individual effects to model unobserved individual heterogeneity.  The model is as follows: \eqref{determinsitic_part} is:
\begin{equation}\label{individual effects}
V_{ijt}=\alpha_j + X_{it}^\top \beta_j + Y_{ijt}^\top \gamma_j + Z_{ijt}^\top \delta,
\end{equation}
where $t=1,\ldots,T_i$ denotes the choice occasion.  In the literature, equation \eqref{individual effects} is often called mixed logit model that permits parameters to vary randomly over commuters. This is done by assuming some continuous heterogeneity distribution (see equation \eqref{instrument}) while keeping the \ref{multinomial} assumption that the error term is i.i.d. Extreme Value type I \citep{Sarrias2016}.

\begin{table}[!h] \centering 
  \caption{Parameter estimates for mode choice with individual effects, walking vs transit vs metro} 
  \label{tab: panel data} 
\begin{tabular}{@{\extracolsep{5pt}}lD{.}{.}{-3} D{.}{.}{-3} } 
\\[-1.8ex]\hline 
\hline \\[-1.8ex] 
 & \multicolumn{2}{c}{\textit{Dependent variable:}} \\ 
\cline{2-3} 
\\[-1.8ex] & \multicolumn{2}{c}{Walk vs Bus vs Metro} \\ 
\\[-1.8ex] & \multicolumn{1}{c}{(1)} & \multicolumn{1}{c}{(2)}\\ 
\hline \\[-1.8ex] 
Intercept$\times$bus  & -0.768^{***} & -0.759^{***} \\ 
  & (0.103) & (0.158) \\ 
\hspace{35pt} $\times$metro & -2.307^{***} & -2.101^{***} \\ 
  & (0.105) & (0.211) \\ 
  Rel. walking distance & -3.253^{***} & -2.395^{***} \\ 
  & (0.889) & (0.925) \\ 
  Rel. walking distance$\times$Rental price (log) & 0.236 & 0.009 \\ 
  & (0.141) & (0.147) \\ 
  Rel. walking distance$\times$Peak & 0.340^{***} & 0.036^{***} \\ 
  & (0.085) & (0.087) \\ 
  Rel. walking distance$\times$Accessibility (Walking) & -0.0013 & -0.0001 \\ 
  & (0.00003) & (0.00003) \\ 
   Rel. walking distance$\times$Load Centrality (Home) & -9.001^{***} & -8.402^{***} \\   
  & (2.597) & (2.575) \\ 
  Rel. walking distance$\times$Load Centrality (School) & -8.299^{***} & -7.588^{***} \\ 
  & (2.586) & (2.569) \\ 
 $\sigma_{\text{rel. walking distance}}$ & 0.0083 & 0.0083 \\ 
  & (0.197) & (0.203) \\ 
  School type controls & \multicolumn{1}{c}{No} & \multicolumn{1}{c}{Yes} \\
 \hline \\
 [-1.8ex] 
Observations & \multicolumn{1}{c}{2,959} & \multicolumn{1}{c}{2,959} \\ 
Log Likelihood & \multicolumn{1}{c}{-1,875.442} & \multicolumn{1}{c}{-1,844.974} \\ 
\hline 
\hline \\[-1.8ex] 
\textit{Note:}  & \multicolumn{2}{r}{$^{*}$p$<$0.1; $^{**}$p$<$0.05; $^{***}$p$<$0.01} \\ 
\end{tabular} 
\end{table}

If it is assumed that the coefficient of relative walking distance varies across individuals according to household income (average residential rental price), departure time (peak) and walking accessibility (walkability and home/school-load centrality) one has

\begin{equation}\label{instrument}
\begin{aligned}
\beta_{\text{rel. walking distance}}= & \beta + \pi_1 (\text{Rental price (log)}) + \pi_2 (\text{Peak}) + \pi_3 (\text{Accessibility (Walking)}) + \\
&\pi_4 (\text{Load Centrality (Home)}) + \pi_5 (\text{Load Centrality (School)}) + \\ 
&\sigma_{\text{rel. walking distance}} \eta_{i},
\end{aligned}
\end{equation}
where $\eta_i \sim N(0,1)$  is the individual-specific scale of the idiosyncratic error term, and $\sigma_{\text{rel. walking distance}}$ is the standard deviation relative to the mean coefficient. Next, one needs to estimate the new $\pi$ coefficients that model the unobserved heterogeneity.   

It can be seen from the results in Table \ref{tab: panel data} that on average the magnitude of the estimated coefficient of relative walking distance is similar to the binary logit model (Table \ref{tab: logit}). However, the standard deviation of relative walking distance, $\sigma_{\text{rel. walking distance}}$ is not significant, therefore, model \eqref{individual effects} under the assumption of  coefficient $\beta_{\text{rel. walking distance}}$ according to  equation \eqref{instrument} is not supported. In other words, to assume unconditional unobserved heterogeneity for relative walking distance cannot be supported. In conclusion, Table \ref{tab: panel data} implies that relative walking distance (so far the main explanatory variable according to Tables \ref{tab: logit} and \ref{tab: mnl and nl}) is not heterogeneous, and the results from the traditional \ref{multinomial} estimator in Table \ref{tab: mnl and nl} are appropriate to use. In fact, different specifications of the coefficient $\beta_{\text{rel. walking distance}}$ were tested, however none of them showed a significant $\sigma_{\text{rel. walking distance}}$ coefficient.  This is probably because of the unbalanced structure of the data. Out of the 2,959 observations, the dataset contains information from 1,832 unique commuters, with an average of 1.6 repeated choices by commuter.

\section{Concluding remarks}\label{sec: conclusions}

This study analyzes the preferences of commuters for short-distance trips. By using granular data from a large-scale experiment in Singapore, the \emph{National Science Experiment 2016}, commuters' (students') behavior was captured during morning trips (home-school). A total of 2,959 trips made by 1,832 different commuters were analyzed and processed to obtain meaningful information about their chosen transportation mode. Tips made by walk bus and metro are analyzed in detail by considering the exact departure time and precise latitude/longitude origin-destination pairs. A substantial difference compared with previous studies \citep{Mackett2003,Limtanakool2006,Guo2009,Vij2013,Ermagun2017} is that commuters in the sample did not directly report any kind information. By exploiting the advantages of Singapore's smart transportation system, a number of variables were computed as proxy attributes for the different transportation modes.

The central questions of this work are: (i) whether or not commuters perceive differences between public transport modes; and (ii) which are the factors that influence commuter's behavior to choose public transport over walking for suitable walking trips. Regarding to the first question, commonly used discrete choice models  are implemented to test for the independence of irrelevant alternatives assumption. Under such assumption, the ratio of, for example, walk and bus stays constant no matter what happens to metro attributes. Surprisingly, the statistical evidence, carried out thought likelihood ratio tests and inclusive value tests, supports the preference of a multinomial structure instead of a nested structure. This is, commuters in Singapore do not differentiate between public transport choices. With respect to the second central question, the calibrated models show that departing during peak-off hours and higher frequency are significant factors that influence their decisions on selecting public transport modes. Conversely, the suggested proxies for walking accessibility, namely land use diversity, $\epsilon-$shortest path algorithm and home-load centrality, are found to be positive correlated with public transport modes. A possible explanation is that some of the metrics were computed using the road street network shapefiles of the city as well as queries from Google Directions API, implying  that the \emph{true} walking paths are not able to be computed. 

The document also provides potential school- and transport-policy implications on enhancing the utilization rates of walking paths. It is found a robust effect of school connectivity on the commuter's decisions to choose the walk mode. In other words, well-connected schools, in terms of road street network connectivity, are more likely to increase the probability of students to decide to walk. It is also found that metro services in the city are as good as bus services due to the non-nested structure of the student's choices. The empirical evidence from the mixed logit model, under a panel data structure, suggests that there is no heterogeneity in relative walking distance among students. The finding implies that transport mode decisions are made purely based on transport mode attributes rather than individual preferences. Therefore, an increase in the quality or efficiency levels  of public transport may be accompanied from  generalized utility gains for all students in the city.

Finally, the limitations of the study need to be highlighted. Given the nature of the experiment, detailed personal characteristics such as gender, age, household income, ethnicity, etc., were not possible to be added into the models. Despite the panel data specification allows to account for individual heterogeneity, the limited repeated choice occasions in the dataset dilute the statistical power of the specification. Additionally, there are suggested the following research directions for future study to improve the model findings: (i) test different entropy indices for walkability as well as different proxies for household income; (ii)  include environmental data (e.g., outdoors temperature, raining fall, etc.) to control for unobserved factors that influence the decision of choosing public transport over walk; and (iii) improve the walking path searching algorithm to account for unobserved walking paths (i.e., short-cuts).



\section*{Acknowledgements}

The research leading to these results is supported by funding
from the National Research Foundation, Prime Minister’s
Office, Singapore, under its Grant RGNRF1402. The first author  would  like  to  acknowledge  CONACyT  
CVU 369933 (Mexico). 

\newpage
\section*{Appendix}
\renewcommand\thetable{A.\arabic{table}}  
\setcounter{table}{0}  

\subsection*{Walk-specific attributes}

For every Subzone $a=1,\ldots,323$,  the entropy index is computed as follows \citep{Frank2005}
\[
\text{Entropy}_a = -\sum_{l=1}^{L_a} \left( \dfrac{d_{al}}{\sum_{l^\prime =1}^{L_a}} \cdot \text{ln}\left(\dfrac{d_{al}}{\sum_{l^\prime =1}^{L_a}} \right) \right) \cdot \dfrac{1}{\text{ln}(L_a)},
\]
where $L_a$ is the number of land use categories and $d$ is the area in square meters. 

Walkability measures are computed using the road network shapefile. However, to obtain such metrics is not a straightforward task. First, a simplifying procedure of the shapefile is needed. The simplifying process helps to: (i) remove redundant nodes and edges that are not needed for the topological structure of a network; (ii) reduce computation time dramatically; and (iii) obtain exact solutions instead of approximate solutions from  heuristic methods.  The following notation and terminology are  useful to formulate the walkability measures introduced in the commuters' utility function. 

Recall that commuters are represented by $i \in I$ and transportation modes by $j \in J$. Next, let the road street network be represented as an undirected \emph{edge-weighted graph} $(\G,w)$ where $\G=(\N, \E)$ is a graph and $w: \E \to \R$ is a weight function (i.e., distance). Here  $\N$ is the set of nodes (with $|\N| = Q$) and $\E$ is the set of edges (with $|\E| = R$). An edge is represented as a pair $(u,v) \in \E$ with $u,v \in \N$. Moreover, $\N(\G)$ and $\E(\G)$ represent the nodes and edges of $\G$, respectively. A node $u$ is an \emph{intermediate node} if there exist two nodes $u^\prime$ and $u^{\prime \prime}$ such that are not direct neighbors themselves. Let $\I \in \N(\G)$ be the set of all intermediate nodes and $I^c \in \N(\G)$ the set of non-intermediate nodes. If there is an edge $(uv) \in \E (\G)$ with endpoints $u$ and $v$, then $u $ and $v$ are adjacent, and $(uv)$ is incident with $u$ and $v$. For each node $u \in \N(\G)$, let $\text{deg}(u)$ denote the degree of $u$ (i.e., the number of edges incident with $u$). A path $\Path (uk)$ is a sequence of nodes $(u,v,\ldots,k)$ with $k\geq 2$ such that $\{ (u,v), \cdots, (k-1,k) \} \in \E(\G)$, where $u$ is called the initial node of $\Path (uk)$ and $k$ the last node of $\Path (uk)$. The distance between two nodes $u,u^\prime \in \Path (uk)$, denoted by $\text{dist}_{\G,w}(u,u^\prime)$, is the length of a shortest $(u,u^\prime)$-path. Given an edge $(u,v) \in \Path (uk)$, $u$ is called a \emph{predecessor} of $v$ and $v$ a \emph{successor} of $u$. Consider the graph $\Hgraph$ with nodes set $\N(\Hgraph)$ and edges set $\E(\Hgraph)$. Then $\Hgraph$ is a subgraph of $\G$ if $\N(\Hgraph) \subseteq \N(\G)$ and $\E(\Hgraph) \subseteq \E(\G)$.

Algorithm 1 describes the steps needed to simplify any road network (lines) shapefile and make it computationally operational. Let denote the pair home-school locations assigned to the nearest node in the road street network by $(o^u,d^*)$, and all existing $K$ walking paths between them as $\Path_\kappa(o^u,d^*), \kappa=1,\dots,K$. The path whose total walking distance is minimum is denoted by $\Path_*(o^u,d^*)$, with $\text{dist}_{\Hgraph,w^h} (o^u,d^*)$ for some $(\Hgraph, w^h ) \subseteq (\G^\prime,w)$. Then, the following Algorithm 2 finds all  paths $\Path_\kappa(o^u,d^*)$ such that they are at most $\epsilon$ times longer than the shortest path $\Path_*(o^u,d^*)$. This is a measure of walking accessibility  due to it allows commuters to find multiple paths under a $\epsilon$ tolerance. During the empirical implementation, $\epsilon=1.1$ implying a 10\% tolerance deviation from the shortest path. Moreover, due to computational time limitations, the counter for different walking paths $c^i$ was limited up to 3,000 ($\delta=3,000$). Algorithms 1-2 are under active development and can be download from \url{https://github.com/Garvit244/Shapefile_to_Network}.

Finally, Algorithm 3 describes the entire procedure of Subzones accessibility. The objective of the procedure is to count how many different Subzones one can reach in a single-trip without transfers while using public transport modes if living in a fixed Subzone. For instance, commuters living in downtown are expected to have higher connectivity than those living in peripheral urban areas.

\begin{table}[h!]
\centering
\caption{Algorithm 1, simplification of the road network graph} 
\label{tab: algorithm cleaning graph}
\begin{tabular}{l} \hline
\textbf{Algorithm 1} (Simplification of the road network graph)                                                                                   \\ \hline
\textbf{Required}: Shapefile with the network graph                                                               \\
\textbf{BEGIN} \\
\qquad \textbf{for all} $u \in \N(\G)$ \textbf{do}                                                                         \\
\qquad \qquad $X^\text{successor}_u:= \{(u_1,u_2,\ldots,u_{k-1},u_k) | u_\ell \text{ are successor nodes of } u \}$ \\ \\
\qquad \qquad \textbf{for all} $u_\ell \in X^\text{successor}_u$ \textbf{do}                                               \\
\qquad \qquad \qquad \textbf{if }$u_\ell \in \I$ \textbf{then}                                                    \\
\qquad \qquad \qquad \qquad $\Path_i(uk) := \{ u,u_1,\ldots,u_s,k \mid u,k \in \I^c, u_\ell \in \I\}$                 \\
\qquad \qquad \qquad \qquad $\Path^u := \bigcup_{u \in \N(\G)} \Path_u(ik)$                                       \\
\qquad \qquad \qquad \textbf{end if} \\ \\
\qquad \textbf{for all} $\Path_u (uk) \in \Path^u$ \textbf{do}                                                             \\
\qquad \qquad $\Path_u^\I (uk) := \Path_u (uk) \backslash \{u,k\}$                                                        \\
\qquad \qquad$ \Path_u^{\I^c} (uk):= \Path_u(uk) \backslash \{u_1,\ldots,u_s\}$                                               \\
\qquad \qquad $w_{uk} = \sum_{(v,v^\prime) \in \Path_u(uk)}\text{dist}_{\G,w}(v,v^\prime)$                                      \\
\qquad \qquad $e_{uk}= (uk;w_{uk})$                                                                              \\ \\
\qquad $\N^\prime:= \N \backslash \bigcup_{u \in \N(\G)} \Path_u^\I (uk)$                                             \\
\qquad $\E^\prime := \{ u,k \in \N^\prime \mid e_{uk}= (uk;w_{uk})\}$                                             \\ \\
\qquad \textbf{Return} $(\G^\prime,w)$                                                                            \\
\textbf{END}                                                                                                       
\\ \hline
\end{tabular}
\end{table}

\begin{table}[h!]
\centering
\caption{Algorithm 2, $\epsilon-$shortest paths}
\label{tab: algorithm epsilon shortest path}
\begin{tabular}{l} \hline
\textbf{Algorithm 2} ($\epsilon-$shortest paths)                                                                                   \\ \hline
\textbf{Required}: $(\G^\prime,w)$, origin and destination sets $\Origin,\Destination$, counter $c^i=0, \theta>0, \epsilon>1, \delta>1.$\\
\textbf{BEGIN} \\
\qquad \textbf{for all} $o^i \in \Origin$ and $d^i \in \Destination$ \textbf{do}                                                                         \\
\qquad \qquad \textbf{Create} a buffer of $\theta$ meters \\
\qquad \qquad \textbf{Extract} subgraphs $(\Hgraph^{o^i},w^{o^i}; \theta) \subseteq (\G^\prime,w)$ and $(\Hgraph^{d^i},w^{d^i}; \theta) \subseteq (\G^\prime,w)$  \\
\qquad \qquad $\N^A:= \N(\Hgraph^{o^i}) \cap \N(\Hgraph^{d^i})$ \\
\qquad \qquad \textbf{Assign} $o^i$ to the nearest node in $\N(\Hgraph^{o^i})$ \\ 
\qquad \qquad $\{o^u \in \N(\Hgraph^{o^i}) \mid \text{dist}_{\Hgraph^{o^i},w^{o^i}} (o^i,o^u) \leq \text{dist}_{\Hgraph^{o^i},w^{o^i}} (o^i,o^\ell) \}$        \\ \\
\qquad \qquad  \textbf{for all} $d^k \in \N^A$ \textbf{do}\\
\qquad \qquad \qquad  \textbf{Get} $\text{dist}_{\Hgraph^{o^i},w^{o^i}}(o^u,d^k)$ \\ \\
\qquad \qquad  $\text{dist}^*_{\Hgraph^{o^i},w^{o^i}} = \underset{d_k}{\text{min}}\{\text{dist}_{\Hgraph^{o^i},w^{o^i}}(o^u,d^k)\}$ \\
\qquad \qquad \textbf{Get} $\Path_\kappa (o^u,d^*), \quad \kappa = 1,\ldots,K$                                                        \\ \\
\qquad \qquad \textbf{for all} $\kappa = 1$ to $K$ \textbf{do}  \\
\qquad \qquad \qquad  \textbf{Get} $\text{dist}^\kappa_{\Hgraph^{o^i},w^{o^i}(o^u,d^*)}$                                      \\ \\
\qquad \qquad   \textbf{if }$\text{dist}^\kappa_{\Hgraph^{o^i},w^{o^i}}(o^u,d^*) \leq \epsilon \cdot \text{dist}^*_{\Hgraph^{o^i},w^{o^i}}(o^u,d^*)$                                      \\
\qquad \qquad \qquad \textbf{Add} one to $c^i$ \\
\qquad \qquad \qquad   \textbf{Stop if} $c^i == \delta$                                             \\
\qquad \qquad \textbf{end if} \\ \\
\qquad \textbf{Return} $c^i$                                                                            \\
\textbf{END}                                                                                                       
\\ \hline
\end{tabular}
\end{table}

\begin{table}[h!]
\centering
\caption{Algorithm 3, Subzone accessibility}
\label{tab: accessibility}
\begin{tabular}{l} \hline
\textbf{Algorithm 3} (Subzone accessibility)\\
\hline
\textbf{Required}: B$_{\text{route}}$(lat, lon, bus, direction, bus\_name), B$_{\text{stop}}$(lat, lon, bus), \\
\qquad \qquad \quad \,\ S(bus, subzone\_id), subzones := $\emptyset$, accessibility\_bus := $\emptyset$, \\
\qquad \qquad \quad \,\  accessibility\_subzone := $\emptyset$	                                     \\
\textbf{BEGIN} 
\\
\qquad \textbf{for all} bus $\in$ B$_{\text{stop}}$ \textbf{do}                \\
\qquad \quad L$^{\text{bus}}$ := \textbf{Get} list of bus\_name for given bus	from B$_{\text{route}}$\\ \\
\qquad \quad \textbf{for all} i $\in$ L$^{\text{bus}}$ \textbf{do}   \\
\qquad \quad \quad direction$_{\text{i}}$ := \textbf{Get} direction of i from B$_{\text{route}}$ \\
\qquad \quad \quad bus\_name$_{\text{i}}$ := \textbf{Get} bus\_name of i from B$_{\text{route}}$ \\
\qquad \quad \quad H$_{\text{bus}^\prime}$ := \{  bus$^\prime$  $\in$ B$_{\text{route}}$ $\mid$   B$_{\text{route}}$\text{(direction)} = direction$_{\text{i}}$ and \\
\qquad \qquad \qquad \qquad \,\  B$_{\text{route}}$\text{(bus\_name)} = bus\_name$_{\text{i}}$\} \\ \\

\qquad \quad \textbf{for all} bus$^\prime$ $\in$ H$_{\text{bus}^\prime}$ \textbf{do} \\
\qquad \quad \quad subzones := \{ subzone\_id $\in$ S \textbar \hspace{0.05cm} S(bus) = bus$^\prime$\} \\ 
\qquad \quad accessibility\_bus[bus] := subzones \\ \\

\qquad S$_{\text{L}}$ := Unique list of subzones from S \\
\qquad \textbf{for all} subzone\_id$^\prime$ $\in$ S$_{\text{L}}$ \textbf{do} \\
\qquad \quad A$_{\text{id}^\prime}$ := \{ bus $\in$ S \textbar \hspace{0.05cm} S(subzone\_id) = subzone\_id$^\prime$ \} \\
\qquad \quad \textbf{for all} id$^\prime$ $\in$ A$_{\text{id}^\prime}$ \textbf{do} \\
\qquad \quad \quad accessibility\_subzone := \{ subzone\_id $\in$ S \textbar \hspace{0.05cm} accessibility\_bus[bus] = id$^\prime$\} \\ \\

\qquad \textbf{Return} accessibility\_subzone\\
\textbf{END} \\
\hline
\end{tabular}
\end{table}

\clearpage
\newpage

\bibliography{mybibfile}

\end{document}